\documentclass[conference]{IEEEtran}

\usepackage{amsmath,amssymb}
\usepackage{mathrsfs}
\usepackage{cite}

\usepackage{graphicx}
\usepackage{booktabs}
\usepackage{array}
\usepackage{url}
\usepackage{tikz}
\usetikzlibrary{arrows.meta,decorations.pathreplacing}

\newcommand{\ind}{\mathbf{1}}
\newcommand{\pospart}[1]{[#1]^+}
\newtheorem{theorem}{Theorem}
\newtheorem{lemma}{Lemma}
\newtheorem{proposition}{Proposition}
\newtheorem{corollary}{Corollary}

\title{Joint Freshness and Deadline Feasibility in Store-and-Forward Satellite IoT Networks.}

\author{\IEEEauthorblockN{Süleyman Çıtır}
\IEEEauthorblockA{Middle East Technical University\\
Ankara, Türkiye\\
suleyman.citir@metu.edu.tr}
}

\begin{document}

\raggedbottom{}

\maketitle

\begin{abstract}
3GPP Release 19 supports store-and-forward satellite access, where traffic accumulated during feeder-link outages must be forwarded through finite, geometry-limited windows. Static priority levels offer little discrimination among large backlogs and ignore the decay in packet value. We formulate QoS feasibility for status updates, measured by destination freshness, and deadline-constrained event reports over periodic two-hop contacts with finite onboard storage. For general calendars, we derive ordering-sensitive miss, cohort-capacity, delivery-rate, and freshness bounds, including an exact status-only characterization. For single status flow and single event class sharing a contact pair per period, a finite-state program gives exact joint feasibility for arbitrary integer age thresholds. Numerical results quantify tradeoffs among load, capacity, deadline, and freshness tolerance.
\end{abstract}


\section{Introduction}
\label{sec:introduction}

Store-and-forward operation allows a satellite to collect traffic while
its feeder link is unavailable and deliver it when ground connectivity
returns. Release~19 supports this mode for satellite access in
E-UTRAN~\cite{3gpp23401}. For satellite IoT, however, accepting a
packet onboard is only the first step toward useful delivery. Its
remaining lifetime must accommodate both the wait for a feeder contact
and competition for the capacity of that contact.

Different applications place different demands on these opportunities.
For routine monitoring, a new status update may replace an older one;
what matters is how current the destination's information remains,
as measured by age of information (AoI)~\cite{yates2021survey}.
An event report, such as a fault alarm, is a distinct message whose
delivery must precede a deadline. Forwarding more status updates can
improve freshness while leaving fewer opportunities for events.
Conversely, postponing a refresh can release capacity for events,
provided that the freshness contract allows it.

Before choosing a scheduler, an operator therefore needs to know
whether the two commitments can be met together. Average capacity
does not settle this question. An event must encounter a service
contact followed by a feeder contact before its deadline, and the
resulting opportunities must also support the required refreshes.
Even separate freshness and event-delivery tests can pass when their
joint requirements are infeasible.

We study feasibility under periodic contacts and finite onboard
storage, with freshness measured by stale-slot fractions and event
service by deadline-miss fractions. Theorem~\ref{prop:floor} gives the
exact status-only stale-slot floor and an attaining policy when resources
suffice. For mixed traffic, Lemma~\ref{lem:joint-reduction} reduces the
specified contact-pair model losslessly, and
Theorem~\ref{thm:joint-characterization} gives an exact finite-program
test for joint feasibility. The phase, cohort-capacity, and aggregate-rate
bounds are supporting necessary tests for general calendars.

\section{Related Work}
\label{sec:related-work}

Predictable contacts are a central resource in disruption-tolerant
satellite networks. Fraire and Finochietto~\cite{fraire2015}
discuss contact-plan design, including the role of routing and traffic
information. More directly, Ar{\i} et al.~\cite{asilomar2025}
study peak-AoI minimization under intermittent satellite connectivity
and store-and-forward delays, comparing different levels of contact
awareness under a transmission budget.

Mixed freshness and deadline-constrained traffic has also been studied
directly. Sun et al.~\cite{sun2021heterogeneous} schedule status updates
alongside timely-throughput traffic over a shared wireless link, seeking
to minimize weighted average AoI subject to timely-throughput
constraints. Fountoulakis et al.~\cite{fountoulakis2023scheduling}
consider an AoI-oriented user and a deadline-constrained user sharing an
error-prone channel; they formulate average-AoI minimization with timely-
throughput constraints as a constrained Markov decision process (CMDP)
and develop dynamic-programming-based policies. Thus, neither the
coexistence of these traffic types nor CMDP modeling is claimed as a
contribution here.

Freshness scheduling over multiple hops is studied by Bedewy
et al.~\cite{bedewy2019multihop}, who establish age-optimality results
for last-generated-first-served policies under specified service
assumptions. Li et al.~\cite{li2023aoi} examine competition between
AoI and service delay in mobile edge caching through achievable-region
bounds and a drift-plus-penalty algorithm. Wang
et al.~\cite{wang2024satelliteaoi} develop statistical peak-AoI,
delay, and error-rate guarantees for satellite--terrestrial networks
in the finite-blocklength regime. These works address related freshness,
delay, or contact effects, but not the joint feasibility problem under
the contact and contract structure considered here.

\section{System Model and QoS Requirements}\label{sec:model}

\subsection{Network and Contact Calendar}

We consider an uplink store-and-forward (S\&F) network of the kind standardized for satellite access in 3GPP Release~19~\cite{3gpp23401}: multiple ground sources, one satellite with onboard storage, and a ground application reached through a gateway. Each packet traverses a source-to-satellite service link and then a satellite-to-ground feeder link, so onboard reception does not constitute delivery.

Time is divided into slots of duration \(\tau\), indexed by \(t\in\mathbb Z_{\ge0}\). The contact calendar repeats every \(P\) slots. All sources share a service window of \(S\) slots per period. The feeder window has duration \(w\) and starts at phase \(\phi\), where

\[
1\le S\le P,\qquad 0\le w\le P,\qquad
\phi\in\{0,\ldots,P-1\}.
\]

All calendar parameters are integers. Choosing the service-window opening as phase zero, the link-availability indicators are

\begin{equation}
\label{eq:indicators}
\begin{aligned}
a(t)&=\ind\{t\bmod P<S\},\\
b_w(t)&=\ind\{(t-\phi)\bmod P<w\}.
\end{aligned}
\end{equation}

The modulo operation takes values in \(\{0,\ldots,P-1\}\), so a feeder window may cross a period boundary. An available service or feeder slot supports at most \(c_{\rm s}\) or \(c_{\rm f}\) equal-sized packets, respectively, both positive integers. The links have separate capacities and operate concurrently: a packet may be collected while a different packet is forwarded in the same slot.

Each hop occupies one slot, which absorbs its transmission and propagation time. A packet collected in slot \(t\) therefore becomes available onboard at \(t+1\) and cannot be forwarded in slot \(t\).

\subsection{Status Updates and Event Reports}

Let \(\mathcal I\) and \(\mathcal J\) denote the finite sets of status flows and event classes, each associated with a source terminal. Every packet \(p\) has a unique identifier, a flow or class label, and a generation time \(g_p\in\mathbb Z_{\ge0}\).

Status flow \(i\in\mathcal I\) generates one update at times \(o_i+n h_i\), \(n\in\mathbb Z_{\ge0}\), where its period \(h_i\ge1\) and offset \(0\le o_i<h_i\) are fixed integers.
Each update contains a complete status observation. An older update may therefore be discarded or replaced and flow is evaluated through destination freshness rather than per-packet delivery~\cite{yates2021survey}.

Event class \(j\in\mathcal J\) generates \(Z_j(t)\) packets at the beginning of slot \(t\), with

\begin{equation}
\label{eq:poisson}
Z_j(t)\sim\operatorname{Poisson}(\lambda p_j),\qquad
p_j>0,\qquad \sum_{j\in\mathcal J}p_j=1.
\end{equation}

The counts are independent across classes and slots. The class proportions \(p_j\) are fixed, while \(\lambda\ge0\) is the aggregate event rate in packets per slot. 

Event packets represent distinct reports and cannot replace one another. Each class has a fixed relative deadline \(D_j\in\mathbb Z_{\ge1}\); packet \(p\) of class \(j\) succeeds only if it reaches the ground application by \(g_p+D_j\), which places the event traffic in the deadline-constrained delivery-ratio setting of~\cite{hou2009theory}. At \(\lambda=0\), no events are generated.

The satellite stores at most \(B\ge1\) waiting packets at slot boundaries. Packets can be discarded at either node. Every discarded event remains a deadline failure in the QoS accounting. An undelivered event is removed once its deadline has been reached. No packet duplication or inter-satellite forwarding is allowed.

\subsection{Scheduling, Buffer Evolution, and Causality}

At boundary \(t\), transmissions from the preceding slot first complete. The system then removes undelivered events whose deadlines have been reached and inserts new source arrivals. Let \(\mathcal U(t)\) and \(\mathcal R(t)\) be the resulting sets of source-held and onboard packets, respectively, before discretionary discards and scheduling. Initial queues are empty.

The controller selects discard sets \(\mathcal D_{\rm s}(t)\subseteq\mathcal U(t)\) and \(\mathcal D_{\rm r}(t)\subseteq\mathcal R(t)\), a collection set \(\mathcal X(t)\), and a forwarding set \(\mathcal Y(t)\). Feasible actions satisfy

\begin{equation}
\label{eq:actions}
\begin{aligned}
\mathcal X(t)&\subseteq
\mathcal U(t)\setminus\mathcal D_{\rm s}(t),\\
|\mathcal X(t)|&\le c_{\rm s}a(t),\\
\mathcal Y(t)&\subseteq
\mathcal R(t)\setminus\mathcal D_{\rm r}(t),\\
|\mathcal Y(t)|&\le c_{\rm f}b_w(t),\\
\big|\mathcal R(t)\setminus
(\mathcal D_{\rm r}(t)\cup\mathcal Y(t))\big|
&+|\mathcal X(t)|\le B.
\end{aligned}
\end{equation}

Let \(\mathcal A(t+1)\) contain all packets generated at \(t+1\), and let \(\mathsf E_{t+1}\) remove undelivered event packets with deadlines at or before \(t+1\). The queue updates are

\begin{equation}
\label{eq:queues}
\begin{aligned}
\mathcal U(t+1)
&=\mathsf E_{t+1}\!\left[
\mathcal U(t)\setminus
(\mathcal D_{\rm s}(t)\cup\mathcal X(t))
\right]\cup\mathcal A(t+1),\\
\mathcal R(t+1)
&=\mathsf E_{t+1}\!\left[
\big(\mathcal R(t)\setminus
(\mathcal D_{\rm r}(t)\cup\mathcal Y(t))\big)
\cup\mathcal X(t)
\right].
\end{aligned}
\end{equation}

Ground deliveries in \(\mathcal Y(t)\) are recorded at \(t+1\), before expiration, so delivery exactly at a deadline is successful. At a common boundary, forwarded packets leave onboard storage before collected packets enter it.

The satellite controller knows the model parameters, contact calendar, status-generation schedules, and initial destination timestamps. At boundary \(t\), it observes the onboard packet identifiers, labels, and generation times in \(\mathcal R(t)\), as well as all ground deliveries completed through \(t\). When \(a(t)=1\), each source reports the metadata of its currently pending packets, including the arrivals at \(t\). Between service contacts, new source arrivals and source-side discards are not directly observed by the satellite. Control messages and  acknowledgments have no additional modeled capacity cost or delay.

Let \(\mathcal H_t^{\rm sat}\) be the information generated by these observations through \(t\), the known parameters, and past satellite actions, and let \(\mathcal H_t^{i}\) be the corresponding local information at the source of flow or class \(i\). Onboard discards and forwarding are \(\mathcal H_t^{\rm sat}\)-measurable; collections are \(\mathcal H_t^{\rm sat}\)-measurable when \(a(t)=1\), with \(\mathcal X(t)=\varnothing\) otherwise; source-side discards are \(\mathcal H_t^{i}\)-measurable. \(\Pi(w)\) denotes the policies respecting these information and action constraints.

In particular, if an event generated at \(g_p\) is collected in slot \(u\) and forwarded in slot \(v\), its successful path must satisfy the ordering and deadline requirements of a valid path,

\begin{equation}
\label{eq:path}
\begin{gathered}
u\ge g_p,\quad a(u)=1,\qquad v\ge u+1,\\
b_w(v)=1,\qquad v+1\le g_p+D_j.
\end{gathered}
\end{equation}

The same ordering applies to status updates. 

\begin{figure}[t]
\centering
\includegraphics[width=\columnwidth]{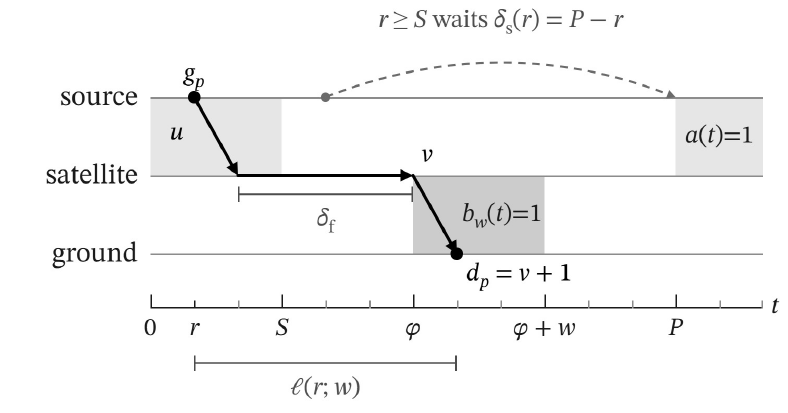}
\caption{Contact calendar and the two-hop path. Service is available in
$[0,S)$ modulo $P$, the feeder window in $[\phi,\phi+w)$. A packet collected
in slot $u$ is onboard at $u+1$ and forwarded no earlier than that, reaching
the ground at $v+1$, so $\ell(r;w)=\delta_{\rm s}(r)+2+\delta_{\rm f}$ as in
\eqref{eq:ell}. Drawn for $P=12$, $S=3$, $\phi=6$, $w=3$, $r=1$: here
$\delta_{\rm s}=0$, $\delta_{\rm f}=4$ and $\ell=6$. A generation at $r\ge S$
finds no service slot left in its period and waits $P-r$.}
\label{fig:calendar}
\end{figure}

\subsection{Freshness and Deadline Metrics}

The ground-application delivery time of packet \(p\) is

\[
d_p:=\inf\{t+1:t\in\mathbb Z_{\ge0},\ p\in\mathcal Y(t)\}.
\]

A discarded or undelivered packet has \(d_p=\infty\). An event received exactly at \(g_p+D_j\) succeeds because deliveries are recorded before expiration at that boundary.

For status flow \(i\), let \(\mathcal P_i\) be its packet set and let \(u_i^0\in\mathbb Z_{\le0}\) be a fixed finite generation timestamp initially known at the destination, which is a reference. The fresh delivered timestamp and destination age, evaluated after all ground deliveries at boundary \(t\), are

\begin{equation}
\label{eq:age}
\begin{aligned}
u_i(t)&=\max\!\left(
\{u_i^0\}\cup
\{g_p:p\in\mathcal P_i,\ d_p\le t\}
\right),\\
\Delta_i(t)&=t-u_i(t).
\end{aligned}
\end{equation}

Consequently,

\begin{equation}
\label{eq:age-rec}
u_i(t+1)=\max\!\left(
\{u_i(t)\}\cup
\{g_p:p\in\mathcal Y(t)\cap\mathcal P_i\}
\right).
\end{equation}

Freshness is evaluated once per slot, immediately after the ground deliveries at its initial boundary. For a fixed integer age threshold \(A_i\ge1\), define the stale-slot fraction

\begin{equation}
\label{eq:stale}
V_i^\pi(\lambda,w)
=\limsup_{T\to\infty,\;T\in\mathbb Z_{>0}}\frac1T
\mathbb E_\pi\!\left[
\sum_{t=0}^{T-1}\ind\{\Delta_i(t)>A_i\}
\right].
\end{equation}

The freshness contract is \(V_i^\pi(\lambda,w)\le\varepsilon_i\), where \(0\le\varepsilon_i<1\) is fixed. Equality \(\Delta_i(t)=A_i\) is fresh, and every slot is counted, including slots without connectivity. Equation~\eqref{eq:stale} is the slot-sampled form of an age-threshold violation metric~\cite{yates2021survey}; it does not integrate the age increase between boundaries.

Let \(N_j(T)=\sum_{t=0}^{T-1}Z_j(t)\) be the number of class-\(j\) events generated before \(T\), and let \(N_j^{\rm miss}(T)\) count the eventual deadline misses among the class-\(j\) events generated in slots \(0,\ldots,T-1\),
\[
N_j^{\rm miss}(T)=\sum_{p\in\mathcal E_j:\,0\le g_p<T}
\ind\{d_p>g_p+D_j\},
\]
where \(\mathcal E_j\) is the set of all generated class-\(j\) events. For \(\lambda>0\), define
\begin{equation}
\label{eq:miss}
M_j^\pi(\lambda,w)
=\limsup_{T\to\infty}
\frac{\mathbb E_\pi\bigl[N_j^{\rm miss}(T)\bigr]}
{\lambda p_j T}.
\end{equation}

The deadline contract is \(M_j^\pi(\lambda,w)\le\eta_j\), where \(0\le\eta_j<1\) is fixed. This is a ratio of expected counts over generation cohorts, not a delay statistic conditioned on successful delivery; discards and buffer overflows are counted as failures. Cohorts are evaluated after their deadlines, even when those deadlines occur after \(T\).

\subsection{Fixed-QoS Traffic-Contact Feasibility}\label{subsec:region}

All QoS targets, status-generation parameters, event proportions, packet and link parameters, \(P,S,\phi\), and \(B\) are fixed. The operating variables are \(\lambda\) and \(w\). We define

\begin{equation}
\label{eq:region}
\begin{aligned}
\mathcal F=\bigl\{&(\lambda,w)\in
\mathbb R_{\ge0}\times\{0,\ldots,P\}:\\
&\exists\pi\in\Pi(w)\quad\text{such that}\\
&V_i^\pi(\lambda,w)\le\varepsilon_i,
\quad\forall i\in\mathcal I,\\
&M_j^\pi(\lambda,w)\le\eta_j,
\quad\forall j\in\mathcal J\bigr\}.
\end{aligned}
\end{equation}

The problem is to derive necessary conditions on \(\mathcal F\), characterize status-only feasibility under sufficient resources, and establish exact joint feasibility in a restricted contact-pair regime.

\section{Traffic-Contact Feasibility}\label{sec:feasibility}

This section develops necessary conditions for the fixed-QoS feasible region in~\eqref{eq:region}. Three mechanisms restrict it. First, the order of the service and feeder windows inside a period decides which arrival phases admit a deadline-feasible path at all, and, for status traffic, exactly how stale the destination must become; on the status-only slice this yields a characterization when both link capacities and onboard storage can support one packet per flow simultaneously. Second, even at phases that admit a path, only a bounded number of packets can traverse the ordered contacts within one deadline. Third, the QoS contracts impose a minimum successful delivery rate, which the link capacities must support. All three apply to every admissible policy, independently of its scheduling rule.

\subsection{Arrival Phase and Earliest Ground Delivery}\label{subsec:phase}

For a packet generated at phase \(r\in\{0,\ldots,P-1\}\), define its waiting time to the first available service slot as

\begin{equation}
\label{eq:dels}
\delta_{\rm s}(r)=
\begin{cases}
0,&r<S,\\
P-r,&r\ge S.
\end{cases}
\end{equation}

For \(w>0\), let \(z(t)=(t-\phi)\bmod P\) and define the waiting time from boundary \(t\) to an available feeder slot by

\begin{equation}
\label{eq:delf}
\delta_{\rm f}(t;w)=
\begin{cases}
0,&z(t)<w,\\
P-z(t),&z(t)\ge w.
\end{cases}
\end{equation}

The earliest possible end-to-end delay of an isolated packet is

\begin{equation}
\label{eq:ell}
\ell(r;w)=\delta_{\rm s}(r)+2+
\delta_{\rm f}\!\left(r+\delta_{\rm s}(r)+1;w\right).
\end{equation}

For \(w=0\), set \(\ell(r;0)=\infty\). The two slots in~\eqref{eq:ell} are the service and feeder transmissions. The feeder waiting time starts after service reception, preserving the requirement that the two hops cannot occur in the same slot.

To obtain~\eqref{eq:ell}, collect the packet in its first available service slot and forward it in the first feeder slot after reception. Because the first feeder slot at or after a boundary is nondecreasing in that boundary, delaying collection cannot yield an earlier feeder opportunity. Thus \(\ell(r;w)\) is the exact isolated-packet delay, with contention removed, and every packet generated at phase \(r\) satisfies \(d_p-g_p\ge\ell(r;w)\) under every policy. An event of class \(j\) has a deadline-feasible contact path if and only if \(\ell(r;w)\le D_j\).

\begin{proposition}[Ordering-Induced Miss Floor]\label{prop:missfloor}
For every \(\lambda>0\) and \(\pi\in\Pi(w)\),
\begin{equation}
\label{eq:q}
\begin{aligned}
M_j^\pi(\lambda,w)&\ge q_j(w),\\
q_j(w)&=\frac1P\sum_{r=0}^{P-1}
\ind\{\ell(r;w)>D_j\}.
\end{aligned}
\end{equation}
Consequently, \(q_j(w)\le\eta_j\) is necessary for feasibility whenever class \(j\) has positive offered traffic.
\end{proposition}

\textbf{Example 1.} Let \(P=10\), \(S=w=1\), and \(D_j=11\). With \(\phi=1\), the delays in~\eqref{eq:ell}, ordered by arrival phase, are

\[
(2,11,10,9,8,7,6,5,4,3),
\]

so \(q_j(1)=0\). With \(\phi=0\), they become

\[
(11,20,19,18,17,16,15,14,13,12),
\]

giving \(q_j(1)=0.9\). Both calendars have the same average service and feeder capacities. In the second calendar, a packet collected in the service slot has already missed that period's feeder transmission opportunity. Any requirement \(\eta_j<0.9\) is therefore infeasible for that calendar at every positive arrival rate. The first calendar passes this ordering test; this alone does not establish its feasibility.

\subsection{A Temporal Capacity Bound for Arrival Cohorts}

The existence of an individual path does not imply that all packets generated at the same phase can use it. Consider the class-\(j\) cohort generated in slot \(r\) of a reference period. All its packets have deadline \(r+D_j\). For any integer \(k\in\{r,\ldots,r+D_j-1\}\), divide its successful packets into those collected before slot \(k\) and those collected in slot \(k\) or later. The first group uses service capacity before \(k\); the second cannot use feeder capacity before \(k+1\). Hence, its successful packet count is at most

\begin{equation}
\label{eq:cut}
\begin{aligned}
C_j(r;w)=
\min_{k=r,\ldots,r+D_j-1}
\Bigl[\;&c_{\rm s}\sum_{u=r}^{k-1}a(u)\\
+\;&c_{\rm f}\!\!\sum_{v=k+1}^{r+D_j-1}\!\!b_w(v)
\Bigr],
\end{aligned}
\end{equation}

where an empty sum is zero. The indicators are evaluated at their actual integer arguments, so~\eqref{eq:cut} also covers deadlines spanning multiple periods.

Equation~\eqref{eq:cut} is a supply bound for the cohort, in the sense used for periodic resource models in real-time scheduling~\cite{shin2003periodic}, with one difference that carries the result: the supply is split across two \emph{ordered} stages rather than drawn from a single resource. Each \(k\) forces every successful packet to consume either service capacity strictly on its earlier side or feeder capacity strictly on its later side, so the two budgets inside a deadline interval cannot be pooled, and minimizing over \(k\) selects the cut separating them most severely. This is what distinguishes calendars differing only in the relative position of their windows. The bound grants the cohort all link capacity and ignores buffer limits and competing packets, so it is an upper bound on its attainable successful count.

For an integer \(C\ge0\), define the expected Poisson excess

\begin{equation}
\label{eq:excess}
H_C(\mu)=\mathbb E\bigl[\pospart{Z-C}\bigr],
\qquad Z\sim\operatorname{Poisson}(\mu).
\end{equation}

This is the first-order loss function of the Poisson distribution, with
\(H_C(\mu)=\mu\Pr\{Z\ge C\}-C\Pr\{Z\ge C+1\}\) and \(H_0(\mu)=\mu\).

\begin{proposition}[Ordering and Cohort-Capacity Bound]\label{prop:cohort}
For every \(\lambda>0\) and \(\pi\in\Pi(w)\),
\begin{equation}
\label{eq:psi}
M_j^\pi(\lambda,w)\ge
\Psi_j(\lambda,w)
\equiv
\frac{1}{\lambda p_j P}
\sum_{r=0}^{P-1}
H_{C_j(r;w)}(\lambda p_j).
\end{equation}
\end{proposition}

If no ordered path exists, choose \(k\) as the earliest service slot in \([r,r+D_j-2]\); the service sum is zero, and the feeder sum must also be zero. If that interval has no service slot, choose \(k=r+D_j-1\). Thus an excluded phase has \(C_j(r;w)=0\). Conversely, an existing ordered path makes every cut in~\eqref{eq:cut} positive, so \(C_j(r;w)\ge1\). Consequently,

\begin{equation}
\label{eq:psiq}
\Psi_j(\lambda,w)\ge q_j(w),
\qquad
\lim_{\lambda\downarrow0}\Psi_j(\lambda,w)=q_j(w),
\end{equation}

since \(H_C(\mu)/\mu\to0\) for each fixed \(C\ge1\). Thus~\eqref{eq:psi} extends Proposition~\ref{prop:missfloor} to a traffic-dependent necessary condition, \(\Psi_j(\lambda,w)\le\eta_j\). It accounts for Poisson batches that exceed the ordered contact capacity even at phases admitting an individual path. Finite deadlines and capacities imply a positive miss floor at every positive Poisson arrival rate, so an exact zero-miss target is infeasible under this arrival model.

Proposition~\ref{prop:cohort} treats each single-slot generation cohort separately. A bound over multi-slot cohorts, applying a common cut to all packets generated in a window whose deadline intervals overlap, may yield additional necessary restrictions; it need not be strictly stronger, since consecutive cohorts can use disjoint service and feeder slots. We do not pursue it here: the single-slot version already separates calendars with equal average capacity, and the aggregate condition of Section~\ref{subsec:rate} captures the coarse effect of sustained competition.

\subsection{Ordering-Induced Freshness Conditions}\label{subsec:fresh}

The ordering restriction expressed by~\eqref{eq:ell} applies to status updates as well, where it limits not the existence of a deadline path but the minimum age at which information can arrive. For status flow \(i\), let

\begin{equation}
\label{eq:ellmin}
\begin{aligned}
\mathcal R_i&=\{(o_i+n h_i)\bmod P: n\in\mathbb Z_{\ge0}\},\\
\ell_i^{\min}(w)&=\min_{r\in\mathcal R_i}\ell(r;w).
\end{aligned}
\end{equation}

\begin{proposition}[Freshness Floor and Required Delivery Rate]\label{prop:fresh}
Fix a status flow \(i\). For every \(\lambda\ge0\) and \(\pi\in\Pi(w)\):
\begin{enumerate}
\item every delivered packet \(p\in\mathcal P_i\) satisfies
      \(d_p-g_p\ge\ell_i^{\min}(w)\);
\item if \(A_i<\ell_i^{\min}(w)\), then \(V_i^\pi(\lambda,w)=1\), so the flow
      is infeasible for every \(\varepsilon_i<1\);
\item if \(A_i\ge\ell_i^{\min}(w)\) and \(V_i^\pi(\lambda,w)\le\varepsilon_i\),
      then, with \(K_i(T)\) the number of delivered packets of flow \(i\) with
      \(d_p<T\),
      \begin{equation}
      \label{eq:rho}
      \liminf_{T\to\infty}\frac{\mathbb E_\pi[K_i(T)]}{T}\ge\rho_i(w)
      \equiv\frac{1-\varepsilon_i}
      {A_i-\ell_i^{\min}(w)+1}.
      \end{equation}
\end{enumerate}
\end{proposition}

Requiring a minimum update rate to hold an age threshold is the standard freshness--rate trade-off~\cite{yates2021survey}; \eqref{eq:rho} makes the constant calendar-dependent through \(\ell_i^{\min}(w)\).

Because flow \(i\) generates only one update every \(h_i\) slots, \(K_i(T)\le T/h_i+1\), so \(\rho_i(w)\le1/h_i\) is necessary as well, at each \(w\). The universal two-slot delivery-rate bound is recovered when \(\ell_i^{\min}(w)=2\), attained when some generation phase of the flow permits collection immediately upon generation and forwarding in the following slot.

The bound~\eqref{eq:rho} constrains the required delivery rate. The stale-slot
fraction itself admits an exact expression, obtained by asking which timestamp
the destination can hold at each boundary. Let

\begin{equation}
\label{eq:gstar}
\begin{aligned}
\sigma(x)&=\max\{u\in\mathbb Z_{\ge0}:u\le x,\ a(u)=1\},\\
\mathcal G_i&=\{o_i+nh_i:n\in\mathbb Z_{\ge0}\},\\
g_i^\star(d)&=\max\{g\in\mathcal G_i:g\le\sigma(d-2)\},
\end{aligned}
\end{equation}

and let \(\mathcal D=\{v+1:v\in\mathbb Z_{\ge0},\ b_w(v)=1\}\) be the feeder completion boundaries,
\(D(t)=\max\{d\in\mathcal D:d\le t\}\), and \(d^+=\min\{d'\in\mathcal D:d'>d\}\).
For \(d\in\mathcal D\), \(g_i^\star(d)\) is the freshest timestamp eligible for
delivery at \(d\) with contention removed, and both \(\sigma\) and
\(g_i^\star\) are nondecreasing. Write \(L_i=\operatorname{lcm}(P,h_i)\) and
\begin{equation}
\label{eq:transient}
T_i^\star=o_i+2P.
\end{equation}

\begin{lemma}[Periodicity]\label{lem:per}
Let \(w\ge1\). For every \(t\ge T_i^\star\) the quantities \(D(t)\) and
\(g_i^\star(D(t))\) are defined and \(g_i^\star(D(t))\ge o_i\ge u_i^0\).
Moreover, for every \(d\in\mathcal D\) with \(d\ge T_i^\star\),
\begin{equation}
\label{eq:perid}
g_i^\star(d+L_i)=g_i^\star(d)+L_i,\qquad {(d+L_i)}^+=d^++L_i.
\end{equation}
Consequently the summand of~\eqref{eq:phi} below depends on \(d\) only through
\(d\bmod L_i\), so \(\Phi_i(w)\) does not depend on which \(L_i\)-period of
\(\mathcal D\) is used.
\end{lemma}

\begin{theorem}[Exact Stale-Slot Floor]\label{prop:floor}
Let \(w\ge1\) and set
\begin{equation}
\label{eq:phi}
\Phi_i(w)=\frac{1}{L_i}\!\!\sum_{d\in\mathcal D_i}\!\!
\pospart{\,d^{+}\!-1-\max\{d-1,\ g_i^\star(d)+A_i\}\,},
\end{equation}
where \(\mathcal D_i\) is any one \(L_i\)-period of
\(\mathcal D\cap[T_i^\star,\infty)\). Then, for every \(\lambda\ge0\) and every 
\(\pi\in\Pi(w)\),
\begin{equation}
\label{eq:floorbound}
\begin{gathered}
\text{(i)}\quad\Delta_i(t)\ \ge\ t-g_i^\star(D(t))
\quad\text{for all }t\ge T_i^\star,\\
\text{(ii)}\quad V_i^\pi(\lambda,w)\ \ge\ \Phi_i(w).
\end{gathered}
\end{equation}
\emph{(iii)}\ 
If \(c_{\rm s}\ge|\mathcal I|\), \(c_{\rm f}\ge|\mathcal I|\) and
\(B\ge|\mathcal I|\), equality holds in the second inequality for the policy
that retains only the newest pending source update per flow, collects it in
each service slot if present, and forwards the newest onboard update of each
flow in each feeder slot if present. Before a collection, it discards any older
onboard update of that flow not selected for forwarding. This construction
discards all events, if present; attainment concerns freshness only.
\end{theorem}

Part~{(iii)} specializes a known fact: serving the most recently generated packet is age-optimal in multihop networks~\cite{bedewy2019multihop}. The contribution is parts~{(i)}--{(ii)}, which evaluate the resulting optimum in closed form for a periodic two-hop calendar with a deterministic generation lattice.

For \(w=0\) no ground delivery occurs and \(V_i^\pi(\lambda,0)=1\), so any
nonempty status-flow set requires \(w\ge1\). A blackout-only lower bound follows by replacing \(g_i^\star(d)\) with its bound
\(d-2\): for the last completion boundary \(d_1\) of a window, \(d_1^+-d_1=P-w+1\)
and the summand of~\eqref{eq:phi} is at least \(\pospart{P-w+2-A_i}\), so

\begin{equation}
\label{eq:gamma}
\Phi_i(w)\ \ge\ \Gamma_i(w)
\equiv\frac{\pospart{P-w+2-A_i}}{P}.
\end{equation}

Equation~\eqref{eq:gamma} is blind to \(\phi\) and to the position of the
service window, which is what \(g_i^\star\) records;~\eqref{eq:phi} is therefore
strictly informative where~\eqref{eq:gamma} is vacuous. As a boundary check,
\(S=P\), \(h_i=1\), \(w=P\) and \(A_i=2\) give
\(\sigma(d-2)=d-2=g_i^\star(d)\), \(d^+=d+1\) and \(\Phi_i(P)=0\), consistent
with \(\ell_i^{\min}(P)=2\) and \(\rho_i(P)=1-\varepsilon_i\).

\textbf{Example 2.} Let \(P=10\), \(S=1\), \(w=1\), \(\phi=0\), \(c_{\rm s}=c_{\rm f}=1\), and one status flow with \(h_i=1\). This is the second calendar of Example~1, whose delays are \((11,20,19,18,17,16,15,14,13,12)\), so \(\ell_i^{\min}(1)=11\): an update collected in the service slot has already missed that period's feeder slot. Two operating points illustrate the gap.

Here \(\Phi_i(1)=1,\,0.9,\,0.8,\,0\) for \(A_i=10,11,12,20\), against
\(\Gamma_i(1)=0.1,\,0,\,0,\,0\). At \(A_i=10\) the flow is infeasible for every
\(\varepsilon_i<1\), whereas~\eqref{eq:gamma} would admit any
\(\varepsilon_i\ge0.1\).

At \(A_i=11\) and \(\varepsilon_i=0.5\) the point is infeasible because
\(\Phi_i(1)=0.9>\varepsilon_i\);~\eqref{eq:rho} detects it independently, since
\(\rho_i(1)=0.5\) exceeds the per-slot delivery capacity
\(\min\{c_{\rm s}S,c_{\rm f}w\}/P=0.1\) of Section~\ref{subsec:rate}, whereas
the universal two-slot form would have given \(0.05\) and admitted it.

\begin{corollary}[Equal Airtime, Unequal Freshness]\label{cor:spread}
Consider one status flow with no event traffic. Fix \(P\ge2\), \(S=w=1\),
\(h_i=1\), \(o_i=0\) and \(A_i=P\). Then
\begin{equation}
\label{eq:spread}
\Phi_i(1;\phi=1)=\frac1P,\qquad
\Phi_i(1;\phi=0)=1,
\end{equation}
while \(\Gamma_i(1)=1/P\) for every \(\phi\). Two calendars with identical
service and feeder airtime therefore differ by \(1-1/P\) in the attainable
stale-slot fraction. The first admits precisely the tolerances
\(\varepsilon_i\in[1/P,1)\), whereas the second admits none with 
\(\varepsilon_i<1\). Every fixed positive tolerance is therefore feasible for
the first calendar for sufficiently large \(P\).
\end{corollary}

We display the dependence on \(\phi\) where it is the object of study;
elsewhere it is suppressed as in Section~\ref{sec:model}.

\subsection{Aggregate Delivery-Rate Condition}\label{subsec:rate}

The successful event rate required by class \(j\) is at least \(\lambda p_j(1-\eta_j)\), and the required status delivery rate is \(\rho_i(w)\) by Proposition~\ref{prop:fresh}. Each delivered packet, of either type, consumes one service transmission and one feeder transmission. Because the two links operate concurrently and have separate capacities, the binding constraint is the smaller of the two per-slot budgets rather than their sum, and

\begin{equation}
\label{eq:cap}
\sum_{i\in\mathcal I}\rho_i(w)
+\lambda\sum_{j\in\mathcal J}p_j\,(1-\eta_j)
\le
\min\!\left\{
\frac{c_{\rm s}S}{P},
\frac{c_{\rm f}w}{P}
\right\}.
\end{equation}

Collected packets that are later discarded consume service capacity as well, so~\eqref{eq:cap} remains valid as an inequality. To relate the generation-cohort definition in~\eqref{eq:miss} to delivery counts strictly before a horizon \(T\), only events generated in the final \(\max_j D_j\) slots can have an on-time delivery at or after \(T\). Their expected count is bounded independently of \(T\), so the difference vanishes after normalization. Initial information likewise contributes only a finite number of fresh boundaries.

\subsection{Outer Region and Scope of the Characterization}

For the mixed-traffic setting, let \(\mathcal O\) consist of the points \((\lambda,w)\), with \(\lambda\ge0\) and \(w\in\{1,\ldots,P\}\), satisfying

\begin{equation}
\label{eq:outer}
\begin{gathered}
\rho_i(w)\le1/h_i,\\[2pt]
\Phi_i(w)\le\varepsilon_i,
\qquad\forall i\in\mathcal I,\\[2pt]
\sum_i\rho_i(w)+\lambda\sum_j p_j\,(1-\eta_j)
\le\min\{c_{\rm s}S/P,c_{\rm f}w/P\},\\[2pt]
\Psi_j(\lambda,w)\le\eta_j
\qquad\forall j\in\mathcal J,\quad\lambda>0,
\end{gathered}
\end{equation}

When \(\mathcal I=\varnothing\), additionally include \((0,0)\) in \(\mathcal O\).
The event inequalities are omitted at \(\lambda=0\). The preceding arguments establish

\begin{equation}
\label{eq:subset}
\mathcal F\subseteq\mathcal O.
\end{equation}

On the status-only slice the inclusion is an equality under the resource
assumptions below. This is a consequence of Theorem~\ref{prop:floor} rather
than an independent result: necessity is part~(ii) and sufficiency is
part~(iii).

\begin{corollary}[A Tight Regime]\label{prop:tight}
Let \(c_{\rm s}\ge|\mathcal I|\), \(c_{\rm f}\ge|\mathcal I|\) and
\(B\ge|\mathcal I|\). Then, for \(w\ge1\),
\begin{equation}
\label{eq:tight}
(0,w)\in\mathcal F
\iff
\Phi_i(w)\le\varepsilon_i\quad\forall i\in\mathcal I ,
\end{equation}
and consequently \(\mathcal F\) and \(\mathcal O\) have the same
\(\lambda=0\) slice.
\end{corollary}

Three of these conditions depend on the position of the windows and not only on their durations: the temporal-cut term in~\eqref{eq:psi}, which couples that dependence to the offered event rate, the minimum-delay term \(\ell_i^{\min}(w)\) in~\eqref{eq:rho}, which couples it to the freshness targets, and \(\Phi_i(w)\) itself, which by~\eqref{eq:spread} separates calendars of equal airtime. The remaining conditions account for the unavoidable feeder gap and the aggregate capacity required by the contracts.

Away from that slice the conditions are not sufficient. The cohort bounds grant overlapping cohorts the same resources; the delivery-rate condition ignores the ordering constraints that~\eqref{eq:cut} makes explicit; and neither enforces the joint competition of status updates and events for the finite onboard buffer, nor the contention among status flows when \(c_{\rm f}<|\mathcal I|\). Points outside \(\mathcal O\) are infeasible; membership in \(\mathcal O\) alone does not establish joint feasibility. Section~\ref{sec:joint-feasibility} gives an exact joint characterization for one status flow and one event class with a single adjacent contact pair per period, synchronized status generation, and deadlines allowing only one eligible contact pair.

\section{Joint Feasibility for Arbitrary Freshness Thresholds}
\label{sec:joint-feasibility}

This section characterizes joint freshness and event-delivery feasibility
for one status flow and one event class with a single adjacent service--feeder
contact pair per period.

\subsection{Contact-Pair Regime and a Lossless Policy Reduction}
\label{subsec:joint-regime}

Consider the dynamics, information constraints, and contracts of
Section~\ref{sec:model}, with one status flow and one event class and specialize
the remaining parameters to
\begin{equation}
\label{eq:joint-regime}
\begin{gathered}
P\ge2,\qquad S=w=1,\qquad\phi=1,\\
h=P,\quad o=0,\quad A\in\mathbb Z_{\ge1},\quad
D\in\{2,\ldots,P+1\}.
\end{gathered}
\end{equation}

Write
\begin{equation}
\label{eq:joint-cycle}
\begin{aligned}
u_k&=kP,\quad v_k=kP+1,\quad d_k=kP+2,\\
\mathcal B_k&=\{d_k,\ldots,d_k+P-1\},\\
C&=\min\{c_{\rm s},c_{\rm f},B\},\qquad m=D-1,\\
\nu&=m\lambda,\qquad q=1-m/P.
\end{aligned}
\end{equation}
Arrivals remain independent Poisson counts of mean $\lambda$ per slot.
The generation cohort $\{(k-1)P+1,\ldots,kP\}$ first encounters service
at $u_k$. Its eligible arrivals are exactly
\begin{equation}
\label{eq:joint-cohort}
N_k=\sum_{t=kP+2-D}^{kP}Z(t)
\overset{\rm iid}{\sim}\operatorname{Poisson}(\nu).
\end{equation}
They can succeed only at $d_k$, since any event in the cohort has deadline
at most $kP+P+1<d_{k+1}$. The fraction $q$ of generation phases is excluded.

An older onboard update can remain useful when $A$ is large. Eliminating
such packets therefore requires a dominance argument.

\begin{lemma}[Lossless Reduction to Current Updates]
\label{lem:joint-reduction}
For feasibility and maximum event-delivery rate under \eqref{eq:joint-regime},
there is no loss in restricting policies to collect and forward, in each
pair $(u_k,v_k)$, at most the current status update and eligible events,
and to retain no packets onboard after that pair. With $X_k$ indicating
delivery of the update generated at $kP$ and $Y_k$ counting event successes,
the reduced actions satisfy
\begin{equation}
\label{eq:joint-batch}
X_k\in\{0,1\},\qquad Y_k=\min\{N_k,C-X_k\}.
\end{equation}
Furthermore, one may always take $X_k=1$ when $N_k<C$, without reducing
event delivery or worsening freshness.
\end{lemma}

\subsection{Exact Age Accounting with a Finite State}
\label{subsec:joint-age-state}

For the reduced policy class, after a current update is first delivered,
let $\ell_k$ be the number of
cycles since the newest delivered generation, measured just after $d_k$.
Then
\begin{equation}
\label{eq:joint-lag}
\begin{gathered}
\ell_k=\begin{cases}0,&X_k=1,\\\ell_{k-1}+1,&X_k=0,\end{cases}
\\
\Delta(d_k+s)=P\ell_k+2+s
\quad(0\le s<P).
\end{gathered}
\end{equation}
Consequently the stale fraction of block $\mathcal B_k$ is
\begin{equation}
\label{eq:joint-age-cost}
\omega_A(\ell_k)
=\frac1P\min\{P,[P(\ell_k+1)+1-A]^+\}.
\end{equation}
For $A=P+1+\kappa$ with $\kappa\ge0$, a delivery followed by one
omitted cycle gives stale fraction $[P-\kappa]^+/P$ in that cycle.

The minimum possible stale fraction is
\begin{equation}
\label{eq:joint-floor-A}
\Phi_A=\omega_A(0)=\frac{[P+1-A]^+}{P}.
\end{equation}
Thus $A=1$ is infeasible for every allowed $\varepsilon<1$. For $A\ge2$, set
\begin{equation}
\label{eq:joint-saturation}
L=\left\lceil\frac{A-1}{P}\right\rceil,\qquad
S_k=\min\{\ell_{k-1},L\}\in\{0,\ldots,L\}.
\end{equation}
All lags at least $L$ have block cost one. Their exact magnitude is irrelevant:
a delivery resets the lag, and an omission keeps the block fully stale. Hence
\begin{equation}
\label{eq:joint-state-transition}
S_{k+1}=\begin{cases}
0,&X_k=1,\\
\min\{S_k+1,L\},&X_k=0.
\end{cases}
\end{equation}
$S_k$ is known before observing $N_k$, and $N_k$ is independent of its history.
The current block has cost $\omega_A(S_{k+1})$.

For any reduced policy, boundary and generation-cohort accounting give
\begin{equation}
\label{eq:joint-accounting}
\begin{aligned}
V^\pi(\lambda,1)
&=\limsup_{n\to\infty}\frac1n
\sum_{k=1}^n\mathbb E_\pi\omega_A(S_{k+1}),\\
M^\pi(\lambda,1)
&=1-\frac1{\lambda P}\liminf_{n\to\infty}\frac1n
\sum_{k=1}^n\mathbb E_\pi Y_k,\qquad\lambda>0.
\end{aligned}
\end{equation}
Initial information and incomplete terminal blocks contribute vanishing
normalized terms. The age formula can be initialized by one current delivery;
before any such delivery, a sufficiently old initial timestamp belongs to
the saturated state.

\subsection{A General Obstruction and an Exact Linear Program}
\label{subsec:joint-characterization}

For $N\sim\operatorname{Poisson}(\nu)$, define
\begin{equation}
\label{eq:joint-functions}
\begin{aligned}
G_c(\nu)&=\mathbb E\min\{N,c\}
=\sum_{j=1}^{c}\Pr\{N\ge j\},\quad G_0=0,\\
p&=\Pr\{N\ge C\}=G_C(\nu)-G_{C-1}(\nu),\\
r&=\lambda P(1-\eta),\qquad
\beta_A=\frac{P(1-\varepsilon)}{A-1}\quad(A\ge2).
\end{aligned}
\end{equation}
For finite $\lambda>0$, $0<p<1$. The quantity $p$ is the probability that
delivering a status update displaces one otherwise successful event.

\begin{proposition}[Joint Necessary Condition for Arbitrary $A$]
\label{prop:joint-obstruction}
For $A\ge2$ and $\lambda>0$, feasibility under \eqref{eq:joint-regime} requires
\begin{equation}
\label{eq:joint-obstruction}
\varepsilon\ge\Phi_A,\qquad
r\le G_C(\nu),\qquad
r\le G_{C-1}(\nu)+1-\beta_A.
\end{equation}
These inequalities need not be sufficient when $A>P+1$.
\end{proposition}

To capture the complete age history, let $z_s$ be the long-run occupation
of state $s$, and let $\xi_s$ be the long-run fraction of cycles with state
$s$, $N_k\ge C$, and $X_k=1$. Define $\mathcal C_A^*(\nu,\varepsilon;C)$
as the value of the following finite linear program:
\begin{equation}
\label{eq:joint-LP}
\begin{aligned}
\mathop{\rm minimize}_{z,\xi}\quad&\sum_{s=0}^L\xi_s\\
\text{subject to}\quad
&\sum_{s=0}^L z_s=1,\quad z_s\ge0,\\
&0\le\xi_s\le pz_s\quad(0\le s\le L),\\
&z_0=1-p+\sum_{s=0}^L\xi_s,\\
&z_s=pz_{s-1}-\xi_{s-1}\quad(1\le s<L),\\
&z_L=pz_{L-1}-\xi_{L-1}+pz_L-\xi_L,\\
&\sum_{s=0}^L\omega_A(s)z_s\le\varepsilon.
\end{aligned}
\end{equation}
The interior balance equations are omitted when $L=1$. An infeasible program
has value $+\infty$. The objective counts status transmissions that displace
events. All other status transmissions occur when $N_k<C$ and have no event
cost. The state cost may be written at $S_k$ rather than $S_{k+1}$ because
their finite-horizon sums differ only by two bounded endpoint terms.

There are $2(L+1)$ decision variables. The unbounded Poisson arrivals have
not been truncated: their effect is represented exactly by $G_C$ and $p$.

\begin{theorem}[Exact Joint Feasibility for Arbitrary Integer $A$]
\label{thm:joint-characterization}
Assume \eqref{eq:joint-regime}, $A\ge2$, and $\lambda>0$.
Program \eqref{eq:joint-LP} is feasible if and only if
$\varepsilon\ge\Phi_A$. When feasible, the maximum successful event count
per cycle, in the $\liminf$ expected-average sense, is
\begin{equation}
\label{eq:joint-throughput}
T_C^*(\nu,\varepsilon;A)
=G_C(\nu)-\mathcal C_A^*(\nu,\varepsilon;C).
\end{equation}
It is attained by a causal policy in the original model. In particular,
\begin{equation}
\label{eq:joint-exact}
(\lambda,1)\in\mathcal F
\iff
\begin{cases}
\varepsilon\ge\Phi_A,\\
\lambda P(1-\eta)\le T_C^*(m\lambda,\varepsilon;A).
\end{cases}
\end{equation}
The minimum event-miss fraction subject to freshness is
\begin{equation}
\label{eq:joint-miss}
\begin{aligned}
M_A^*(\lambda)
&=1-\frac{T_C^*(m\lambda,\varepsilon;A)}{\lambda P}\\
&=q+\frac{H_C(\nu)+\mathcal C_A^*(\nu,\varepsilon;C)}{\lambda P},
\end{aligned}
\end{equation}
where $H_C(\nu)=\nu-G_C(\nu)$ is the Poisson excess in \eqref{eq:excess}.
At $\lambda=0$, feasibility is equivalent to $\varepsilon\ge\Phi_A$.
\end{theorem}

\subsection{Closed-Form Characterizations}
\label{subsec:joint-specializations}

\begin{corollary}[Thresholds No Larger Than One Fresh Block]
\label{cor:joint-small-A}
Let $2\le A\le P+1$ and $\varepsilon\ge\Phi_A$. Set
\begin{equation}
\label{eq:joint-effective-epsilon}
\delta_A=\frac{\varepsilon-\Phi_A}{1-\Phi_A}=1-\beta_A.
\end{equation}
Then
\begin{equation}
\label{eq:joint-small-A}
T_C^*(\nu,\varepsilon;A)
=\min\{G_C(\nu),G_{C-1}(\nu)+\delta_A\}.
\end{equation}
In particular, at $A=P+1$, $\Phi_A=0$, $\delta_A=\varepsilon$, and
\begin{equation}
\label{eq:joint-Pplus1}
T_C^*(\nu,\varepsilon;P+1)
=\min\{G_C(\nu),G_{C-1}(\nu)+\varepsilon\}.
\end{equation}
\end{corollary}

\begin{corollary}[Zero Stale Fraction with Arbitrary $A$]
\label{cor:joint-zero-stale}
Let $A\ge P+1$, $\varepsilon=0$, and
$R=\lfloor(A-1)/P\rfloor$. Then
\begin{equation}
\label{eq:joint-zero-stale}
\begin{aligned}
\mathcal C_A^*(\nu,0;C)
&=\frac{p^R}{\sum_{j=0}^{R-1}p^j},\\
T_C^*(\nu,0;A)&=G_C(\nu)-\frac{p^R}{\sum_{j=0}^{R-1}p^j}.
\end{aligned}
\end{equation}
An attaining policy refreshes whenever $N_k<C$; when $N_k\ge C$, it
refreshes only if $S_k=R-1$.
\end{corollary}

$R$ is the number of complete fresh blocks that one delivery can support.
At $A=P+1$, $R=1$ and every busy cycle must carry status if $\varepsilon=0$.
At $A=2P+1$, $R=2$: one cycle may be skipped, and the minimum busy-refresh
rate decreases from $p$ to $p^2/(1+p)$.

\subsection{Admission and Provisioning}
\label{subsec:joint-deployment}

Within the regime of \eqref{eq:joint-regime}, the joint characterization
answers three deployment questions: how much traffic to admit, how much
packet capacity to provide, and how much freshness flexibility is needed.
First check $\varepsilon\ge\Phi_A$; otherwise, increasing link capacity
or storage cannot meet the freshness contract. When this check passes,
the required effective capacity is
\begin{equation}
\label{eq:joint-capacity-required}
\begin{aligned}
C_{\rm req}(\lambda;A)=\min\{c\in\mathbb Z_{\ge1}:\;&\lambda P(1-\eta)\\[-2pt]
&\le T_c^*(m\lambda,\varepsilon;A)\},
\end{aligned}
\end{equation}
and feasibility requires and is ensured by
\begin{equation}
\label{eq:joint-provision}
c_{\rm s},\ c_{\rm f},\ B\ \ge C_{\rm req}(\lambda;A).
\end{equation}
For fixed resources, the admission limit and minimum acceptable age
threshold are
\begin{equation}
\label{eq:joint-admission}
\lambda_{\rm adm}(C;A)=\sup\bigl(\{0\}\cup
\{\lambda>0:r\le T_C^*(m\lambda,\varepsilon;A)\}\bigr),
\end{equation}
\begin{equation}
\label{eq:joint-Arequired}
\begin{aligned}
A_{\rm req}=\min\{a\in\mathbb Z_{\ge2}:\;&\varepsilon\ge\Phi_a,\\[-2pt]
&r\le T_C^*(m\lambda,\varepsilon;a)\}.
\end{aligned}
\end{equation}
An empty minimum is $\infty$; each search varies only its stated
parameter while holding the other parameters and targets fixed.

\begin{proposition}[Provisioning and Admission]
\label{prop:joint-provisioning}
For $\lambda>0$ and $\varepsilon\ge\Phi_A$, a finite
$C_{\rm req}$ exists if and only if $\eta>q$. The optimal event rate
is nondecreasing in $C$, and increasing $A$ cannot destroy feasibility.
Feasibility is downward closed in $\lambda$, so a scalar search using
\eqref{eq:joint-LP} determines \eqref{eq:joint-admission}.
\end{proposition}

The distinction is operational: if $\varepsilon<\Phi_A$ or
$\eta\le q$, adding packet capacity cannot repair the deployment.
Otherwise, the characterization supports admission control and integer
searches over capacity or age threshold. Section~\ref{sec:numerical-results} compares these choices for a reference
deployment.

\section{Numerical Results and Deployment Implications}
\label{sec:numerical-results}

The reference deployment
has one status flow and one event class, with $P=h=10$, $S=w=1$,
$\phi=1$, and $o=0$. Unless varied in a figure, the remaining parameters
are $c_{\rm s}=c_{\rm f}=B=2$, $D=11$, $\lambda=0.1$,
$A=11$, $\varepsilon=0$, and $\eta=0.2$.
We write $C=\min\{c_{\rm s},c_{\rm f},B\}$ for the effective
packet capacity of the contact pair.

The curves evaluate the exact joint characterization in MATLAB.
To check its attainment, we also simulate individual packets through
both hops, enforcing onboard storage, deadlines, and post-delivery
age sampling.

\subsection{The Cost of Sharing Contacts}
\label{subsec:numerical-joint-gap}

\begin{figure}[!t]
\centering
\includegraphics[width=0.88\columnwidth]{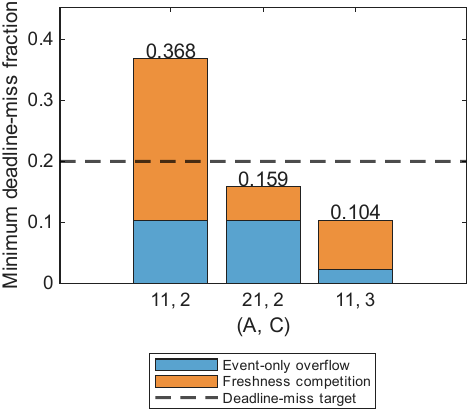}
\caption{Exact event-loss decomposition at $\lambda=0.1$,
$D=11$, and $\varepsilon=0$. The pairs on the horizontal axis
specify $(A,C)$. Contact-timing loss is zero for all three cases;
the dashed line is the deadline-miss target.}
\label{fig:joint-loss-decomposition}
\end{figure}

Figure~\ref{fig:joint-loss-decomposition} explains the gap at the
reference load. Events alone meet the target. Requiring $A=11$ with no stale
slots raises loss to $0.3679$: status refreshes displace event
deliveries even though every arrival phase admits a timely two-hop
path at $D=11$. The timing-loss component is therefore zero in this
example. Increasing $A$ to $21$ reduces the refresh cost sufficiently
to bring the total loss below the target, at $0.1589$. Increasing all
three resources to three achieves $0.1036$ with $A=11$; in that case,
the event-only overflow itself is smaller, but freshness competition
still contributes to the total. The two remedies thus recover
feasibility through different changes in the loss budget.

\subsection{How Much Freshness Flexibility Is Useful?}
\label{subsec:numerical-freshness}

\begin{figure}[!t]
\centering
\includegraphics[width=0.88\columnwidth]{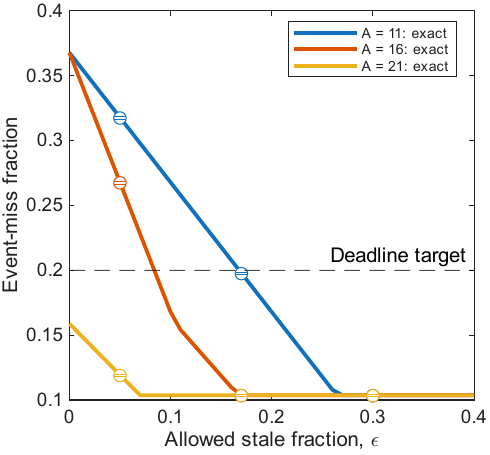}
\caption{Minimum event-miss fraction as the stale allowance varies
for $A\in\{11,16,21\}$. Lines evaluate the exact characterization;
markers and intervals are packet-level simulation estimates.
The dashed line marks $\eta=0.2$.}
\label{fig:freshness-contract-tradeoff}
\end{figure}

A freshness contract can be relaxed by increasing either the age
threshold or the permitted stale fraction. Their effects are related
but distinct. In Fig.~\ref{fig:freshness-contract-tradeoff}, $A=11$
and $A=16$ have the same event loss at $\varepsilon=0$, but allowing
stale slots immediately separates their curves. A missed refresh
produces fewer stale boundaries at the larger threshold, allowing
more capacity to be reassigned to events for the same stale allowance.
At $A=11$, an allowance of approximately $0.17$ is needed to meet
the target. With $A=21$, the target is already met at zero
staleness.

\begin{figure}[!t]
\centering
\includegraphics[width=0.88\columnwidth]{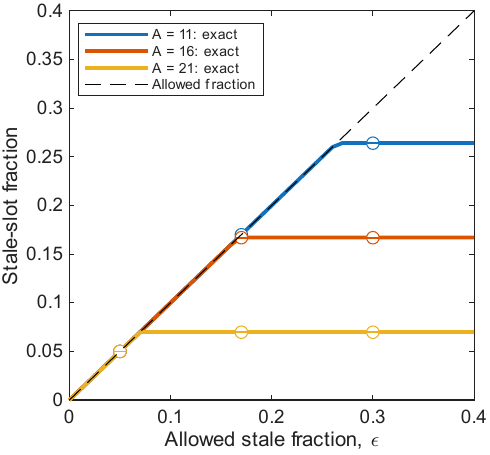}
\caption{Stale fraction of the selected attaining policy under
the contracts in Fig.~\ref{fig:freshness-contract-tradeoff}.
The dashed diagonal is the allowed fraction. Markers and intervals
show packet-level estimates.}
\label{fig:freshness-contract-usage}
\end{figure}

All three curves eventually reach the event-only floor of $0.1036$.
Beyond this point, relaxing freshness further releases no additional
useful event capacity. This saturation also appears in
Fig.~\ref{fig:freshness-contract-usage}: the selected attaining policy
uses the available stale budget until it reaches the event-only
limit, then leaves the additional allowance unused. The close agreement
between the packet estimates and analytical curves holds for both
the event loss and the resulting stale fraction.

\subsection{Using the Freshness Margin}
\label{subsec:numerical-mechanism}

\begin{figure}[!t]
\centering
\includegraphics[width=0.88\columnwidth]{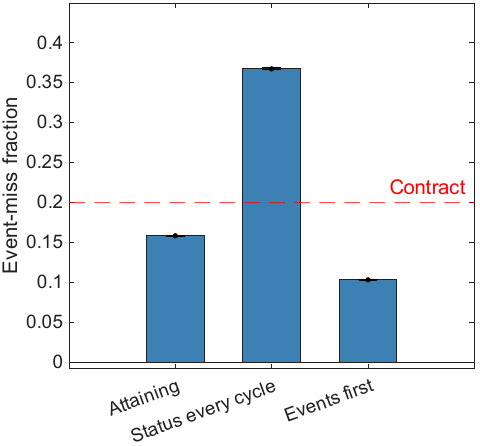}
\caption{Simulated event-miss fractions for three allocation rules
at $A=21$ and $\varepsilon=0$. The dashed line marks $\eta=0.2$;
error bars are pointwise 95\% confidence intervals.}
\label{fig:policy-mechanism}
\end{figure}

At $A=21$, the destination can remain fresh allowing the scheduler to share contact capacity. Figure~\ref{fig:policy-mechanism} compares simulated event-miss fractions for three rules: refreshing every cycle yields $0.37$, event-first service yields $0.1$, and the attaining policy yields $0.16$.
Event-first service achieves the lowest miss fraction but produces a stale fraction of $0.07$, violating the zero-stale contract. The attaining policy instead uses the available freshness margin to defer updates when possible while maintaining the required freshness.

\subsection{From Feasibility to a Deployment Choice}
\label{subsec:numerical-admission}

\begin{figure}[!t]
\centering
\includegraphics[width=0.88\columnwidth]{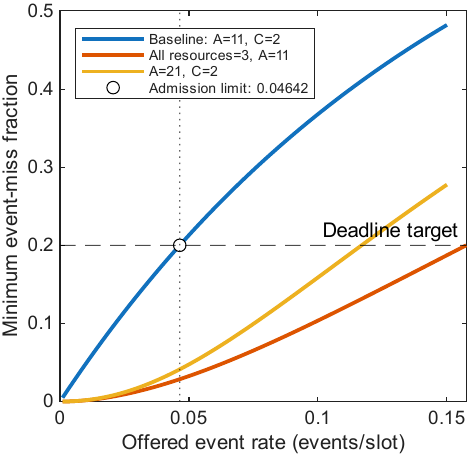}
\caption{Exact event-miss fractions versus offered rate for
$D=11$ and $\varepsilon=0$. The marked crossing gives the baseline
admission limit. The capacity upgrade sets
$c_{\rm s}=c_{\rm f}=B=3$.}
\label{fig:admission-and-deadline}
\end{figure}

Figure~\ref{fig:admission-and-deadline} gives the admission consequence
of these alternatives. With the original resources and $A=11$, the
admissible rate ends at approximately $0.04642$ events per slot.
Supporting $\lambda=0.1$ therefore requires a change: either increase
both link capacities and storage to three, or retain the resources
and accept $A=21$. Increasing the feeder capacity alone leaves
$C=2$ and the baseline curve unchanged, because service reception
and onboard storage remain bottlenecks.

\begin{figure}[!t]
\centering
\includegraphics[width=0.88\columnwidth]{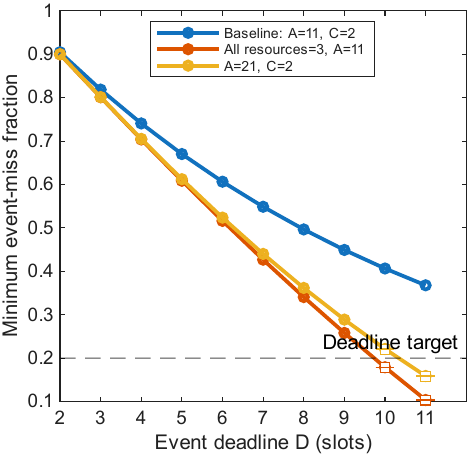}
\caption{Deadline sensitivity at $\lambda=0.1$ and
$\varepsilon=0$. Lines are exact evaluations; square markers and
intervals show packet-level estimates. The dashed line marks
the deadline-miss target.}
\label{fig:deadline-sensitivity}
\end{figure}

The deadline must be included in that decision.
Figure~\ref{fig:deadline-sensitivity} keeps the load at $0.1$ and
varies $D$. At $D=10$, the capacity upgrade gives a minimum miss
fraction of $0.1791$ and meets the contract, whereas relaxing the
age threshold gives $0.2212$ and fails. Both remedies become feasible
at $D=11$. An operator retaining the original hardware can therefore
admit up to $0.04642$ events per slot under the original contracts,
or support $0.1$ by allowing $A=21$ when $D=11$ is acceptable.
If the application instead requires $D=10$, increasing both link
capacities and storage to three supports the same load while keeping
$A=11$. These choices follow from the joint contracts and contact
calendar; an improvement in either freshness tolerance or feeder
capacity alone need not resolve the deployment's limiting constraint.

\section{Conclusion}
\label{sec:conclusion}

The principal finding is that separate freshness and deadline tests, as
well as average capacity alone, do not determine whether both contracts
can be met under ordered two-hop contacts. Theorem~\ref{prop:floor} gives
the exact stale-slot floor for status-only operation and an attaining
policy when resources suffice. For mixed traffic, Lemma~\ref{lem:joint-reduction}
shows that the restricted contact-pair model admits a lossless reduction
to current status updates and eligible events; Theorem~\ref{thm:joint-characterization}
then characterizes joint feasibility by a finite occupation program.

The phase, cohort-capacity, and aggregate-rate bounds support screening
under general calendars but are not sufficient for joint feasibility.
The exact mixed-traffic result is restricted to one status flow and one
Poisson event class, synchronized status generation with $h=P$ and $o=0$, one
adjacent one-slot service--feeder pair per period ($S=w=1$, $\phi=1$),
and deadlines $D\in\{2,\ldots,P+1\}$. Within this regime it covers all
integer age thresholds $A\ge2$ (while $A=1$ is infeasible for the stated
freshness targets). Extending exact joint characterization to general
contact calendars, multiple contact pairs, and multiple flows or event
classes remains open.

\clearpage
\appendices
\section{Proofs for Traffic-Contact Feasibility}

\subsection{Proof of Proposition~\ref{prop:missfloor}}

\begin{IEEEproof}
Put \(\mu_j=\lambda p_j>0\). Every packet generated at an excluded phase misses under every policy. Since \(\mathbb E[Z_j(kP+r)]=\mu_j\),

\[
\begin{aligned}
\mathbb E_\pi[N_j^{\rm miss}(nP)]
&\ge n\mu_j\sum_{r=0}^{P-1}
\ind\{\ell(r;w)>D_j\}\\
&=n\mu_j P q_j(w).
\end{aligned}
\]

Dividing by \(\mathbb E[N_j(nP)]=nP\mu_j\) shows that the ratio in~\eqref{eq:miss} is at least \(q_j(w)\) for every \(n\), hence so is its limit superior. The phase share is \(1/P\) in expected offered counts; realized counts need not be identical across phases.
\end{IEEEproof}

\subsection{Proof of Proposition~\ref{prop:cohort}}

\begin{IEEEproof}
Let \(L_{j,t}\) be the number of eventual deadline misses in the class-\(j\) cohort generated at \(t\). For \(t=kP+r\), the periodic temporal-cut bound~\eqref{eq:cut} limits that cohort's successful deliveries to \(C_j(r;w)\), regardless of other traffic. Hence, on every sample path,

\[
L_{j,kP+r}\ge
\pospart{Z_j(kP+r)-C_j(r;w)}.
\]

The generation cohorts are disjoint, even when their delivery intervals overlap. With \(\mu_j=\lambda p_j\), their expected misses therefore satisfy

\[
\begin{aligned}
\mathbb E_\pi[N_j^{\rm miss}(nP)]
&=\sum_{k=0}^{n-1}\sum_{r=0}^{P-1}
\mathbb E_\pi[L_{j,kP+r}]\\
&\ge n\sum_{r=0}^{P-1}H_{C_j(r;w)}(\mu_j).
\end{aligned}
\]

Divide by \(nP\mu_j\) and use the complete-period subsequence in~\eqref{eq:miss} to obtain~\eqref{eq:psi}. All cohorts are scored after their deadlines, as required by~\eqref{eq:miss}.
\end{IEEEproof}

\subsection{Proof of Proposition~\ref{prop:fresh}}

\begin{IEEEproof}
(i) Each \(g_p\) has a phase in \(\mathcal R_i\), and by Section~\ref{subsec:phase} a packet generated at phase \(r\) cannot be delivered earlier than \(g_p+\ell(r;w)\).

(ii) Boundary \(t\) is fresh only if \(u_i(t)\ge t-A_i\). By (i), any delivered packet with \(d_p\le t\) has \(g_p\le d_p-\ell_i^{\min}(w)\le t-\ell_i^{\min}(w)<t-A_i\), so no delivery can make \(t\) fresh. Only the initial timestamp can, and \(u_i^0\ge t-A_i\) holds for at most finitely many \(t\). The fresh-boundary count is therefore bounded uniformly in \(T\), and~\eqref{eq:stale} equals one.

(iii) A delivered packet \(p\) can make boundary \(t\) fresh only if \(d_p\le t\le g_p+A_i\), so it accounts for at most
\[
\pospart{g_p+A_i-d_p+1}
\le A_i-\ell_i^{\min}(w)+1
\]
fresh boundaries, using (i). Let \(F_i(T)=\sum_{t=0}^{T-1}\ind\{\Delta_i(t)\le A_i\}\). Every fresh boundary \(t<T\) is accounted for by the initial timestamp or by a delivered packet with \(d_p\le t<T\), hence counted in \(K_i(T)\). Therefore
\[
F_i(T)\le b_i^0+\bigl(A_i-\ell_i^{\min}(w)+1\bigr)K_i(T)
\]
for a finite constant \(b_i^0\) accounting for initial information. Since~\eqref{eq:stale} implies \(\liminf_{T\to\infty}\mathbb E_\pi[F_i(T)]/T\ge1-\varepsilon_i\), the claim follows.
\end{IEEEproof}

\subsection{Proof of Lemma~\ref{lem:per}}

\begin{IEEEproof}
Since \(S\ge1\) and \(w\ge1\), every \(P\) consecutive slots contain a service
slot and a feeder slot, so \(\sigma(x)\ge x-P+1\) and \(D(t)\ge t-P+1\) wherever
they are defined. Hence \(\sigma(D(t)-2)\ge t-2P\ge o_i\) for \(t\ge
T_i^\star\), which makes the set defining \(g_i^\star(D(t))\) nonempty, and
\(g_i^\star(D(t))\ge o_i\ge0\ge u_i^0\). For~\eqref{eq:perid}, the indicators
\(a(\cdot)\) and \(b_w(\cdot)\) have period \(P\) and \(P\mid L_i\), so
\(\sigma(x+L_i)=\sigma(x)+L_i\) and \(\mathcal D+L_i=\mathcal D\cap[L_i+1,\infty)\); and
\(h_i\mid L_i\) gives \(\max\{g\in\mathcal G_i:g\le y+L_i\}
=\max\{g\in\mathcal G_i:g\le y\}+L_i\). These identities give both
equalities.
\end{IEEEproof}

\subsection{Proof of Theorem~\ref{prop:floor}}

\begin{IEEEproof}
Let \(t\ge T_i^\star\). A packet \(p\in\mathcal P_i\) with \(d_p\le t\) was forwarded in slot
\(v=d_p-1\), hence was onboard at \(v\) and collected in some slot
\(u\le d_p-2\) with \(a(u)=1\). Therefore
\(g_p\le u\le\sigma(d_p-2)\) and so \(g_p\le g_i^\star(d_p)\). Since
\(d_p\in\mathcal D\) and \(d_p\le t\) give \(d_p\le D(t)\), and \(g_i^\star\) is
nondecreasing, every delivered packet counted in~\eqref{eq:age} satisfies
\(g_p\le g_i^\star(D(t))\), and \(u_i^0\le g_i^\star(D(t))\) by
Lemma~\ref{lem:per}. Hence \(u_i(t)\le g_i^\star(D(t))\), which is part~(i).
Part~(ii) follows next, and Lemma~\ref{lem:per} makes \(\Phi_i(w)\)
well defined.

Fix \(d\in\mathcal D\). For \(t\in\{d,\ldots,d^+-1\}\) we have \(D(t)=d\), so
\(\Delta_i(t)>A_i\) whenever \(t>g_i^\star(d)+A_i\), and the number of such
boundaries in that interval is exactly the summand of~\eqref{eq:phi}. The
intervals \(\{d,\ldots,d^+-1\}\) are disjoint and tile the boundaries after the
transient, and the summand depends on \(d\) only through \(d\bmod L_i\);
averaging over one \(L_i\)-period gives \(V_i^\pi(\lambda,w)\ge\Phi_i(w)\).

For attainment, the stated policy uses at most one transmission per flow on
each link and leaves at most one waiting onboard packet per flow after each
slot. Its discards and selections therefore satisfy~\eqref{eq:actions} under
the stated capacity assumptions and use only locally available information.

Let \(v=D(t)-1\), \(u=\sigma(v-1)\), and \(g=g_i^\star(D(t))\).
The packet generated at \(g\) is the newest update generated by slot \(u\).
If still source-held, it is collected in slot \(u\); otherwise it has already
been collected and is either onboard or delivered. It cannot have been
discarded in favor of a newer collected update before \(v\): no newer update
was generated by \(u\), and no service slot lies between \(u+1\) and \(v-1\).
Thus the packet is either delivered before \(v\), or is onboard and forwarded
in slot \(v\). A collection in slot \(v\) does not discard a packet selected
for forwarding. Hence \(u_i(t)\ge g\), and the preceding upper bound on the
delivered timestamp gives \(u_i(t)=g_i^\star(D(t))\) after the transient.
This attains~\eqref{eq:floorbound} and therefore \(\Phi_i(w)\).
\end{IEEEproof}

\subsection{Proof of Corollary~\ref{prop:tight}}

\begin{IEEEproof}
Necessity is part~(ii) of Theorem~\ref{prop:floor}. For sufficiency, the event
contracts are vacuous at \(\lambda=0\), and the policy in part~(iii) of
Theorem~\ref{prop:floor}
attains \(V_i^\pi(0,w)=\Phi_i(w)\le\varepsilon_i\) for every flow, so
\((0,w)\in\mathcal F\), which gives~\eqref{eq:tight}. For the slices with \(w\ge1\), every
inequality of~\eqref{eq:outer} is necessary for feasibility, so all of them hold
at such a point and \((0,w)\in\mathcal O\); conversely a point of
\(\mathcal O\) with \(\lambda=0\) satisfies \(\Phi_i(w)\le\varepsilon_i\) and is
therefore feasible by~\eqref{eq:tight}. For \(w=0\), both sets contain
\((0,0)\) exactly when \(\mathcal I=\varnothing\).
\end{IEEEproof}

\section{Proofs for Joint Feasibility}

\subsection{Proof of Lemma~\ref{lem:joint-reduction}}

\begin{IEEEproof}
Events generated between consecutive service slots can succeed only through
the following pair: their latest possible deadline is $d_{k+1}-1$.
Thus no successful event can be carried across pairs.

Put $K=\min\{c_{\rm f},B\}$. If $c_{\rm s}\ge K$, retain all its successful
events at $d_k$ and replace any status deliveries at $d_k$ by one delivery
of the current update. Their total count is at most $K$, since all original
forwarded packets were onboard at $v_k$ and obeyed the feeder limit.
The replacement set therefore fits service, feeder, and buffer capacities
when collected at $u_k$. Its timestamp is at least every replaced timestamp,
so destination age cannot increase. The choice is causal: between $u_k$ and
$v_k$ there is no new source observation; any private randomization used at
$v_k$ can be drawn at $u_k$.

If $c_{\rm s}<K$, instead advance every collected status packet to completion
in its collection pair, replacing multiple collected status packets by the
single current update. Forward all eligible events collected in that pair.
There are at most $c_{\rm s}$ such packets, which fit both feeder and buffer
capacities. Every originally successful event still succeeds. Whenever the
emulated policy eventually delivers a status packet, the new policy has
already delivered an update with a timestamp at least as recent in that
packet's collection pair. Its destination age therefore cannot be larger.
Both constructions leave no onboard backlog and need only observed metadata
and a simulated history of the original policy, not future arrivals.

Within the resulting class, serving $\min\{N_k,C-X_k\}$ events can only
increase successful deliveries. If $N_k<C$, adding a current status update
when none was selected leaves this event count unchanged and weakly improves
all future ages; the original future decisions can still be emulated.
Thus refreshing in such cycles also incurs no loss. The observation constraints are preserved.
\end{IEEEproof}

\subsection{Proof of Proposition~\ref{prop:joint-obstruction}}

\begin{IEEEproof}
A delivery at $d_k$ has timestamp $kP$ and sampled age two. Omissions
increment the generation lag by one cycle, giving \eqref{eq:joint-lag}.
Counting integers $s\in\{0,\ldots,P-1\}$ with $P\ell_k+2+s>A$ proves
\eqref{eq:joint-age-cost}. For $\ell\ge L$, the first sample is already
stale, so saturation preserves every block cost and future transition.
The lowest cost is $\omega_A(0)=\Phi_A$.

The blocks tile the boundary samples after startup. Generation cohorts
likewise partition event arrivals; their expected offered count per cycle
is $\lambda P$. Initial and terminal partial cohorts contribute $O(P\lambda)$
expected packets. These facts give \eqref{eq:joint-accounting}, including
its limsup/liminf conventions.

Each delivered current update can account for at most $A-1$ fresh boundaries
starting at its delivery. A union bound, including finite initial information,
therefore gives
\[
\#\{\text{fresh boundaries in $n$ blocks}\}
\le (A-1)\sum_{k=1}^n X_k+O(1).
\]
Freshness implies $\liminf_n n^{-1}\sum_k\mathbb E X_k\ge\beta_A$.
On every reduced sample path,
\[
Y_k\le\min\{N_k,C\},\qquad
Y_k\le\min\{N_k,C-1\}+1-X_k.
\]
Expectations and the event contract yield \eqref{eq:joint-obstruction}.
Lemma~\ref{lem:joint-reduction} makes these necessary for the original
policy class as well.
\end{IEEEproof}

\subsection{Proof of Theorem~\ref{thm:joint-characterization}}

\begin{IEEEproof}
For a reduced policy that always refreshes when $N_k<C$, define expected
empirical measures
\begin{align*}
z_s^{(n)}&=\frac1n\sum_{k=1}^n\Pr\{S_k=s\},\\
\xi_s^{(n)}&=\frac1n\sum_{k=1}^n
\Pr\{S_k=s,N_k\ge C,X_k=1\}.
\end{align*}
Independence of $N_k$ from the prior state gives
$0\le\xi_s^{(n)}\le pz_s^{(n)}$. A visit to $s$ resets with mass
$(1-p)z_s^{(n)}+\xi_s^{(n)}$ and advances with mass
$pz_s^{(n)}-\xi_s^{(n)}$. Finite-horizon flow imbalance is an endpoint
term of order $1/n$. Consequently every subsequential limit of these
bounded measures satisfies the balance and normalization constraints in
\eqref{eq:joint-LP}. The freshness contract ensures its cost constraint.

Moreover, \eqref{eq:joint-batch} gives the exact expected-average identity
\[
\frac1n\sum_{k=1}^n\mathbb E Y_k
=G_C(\nu)-\sum_{s=0}^L\xi_s^{(n)}.
\]
Every feasible limit point has objective at least $\mathcal C_A^*$, so the
event rate cannot exceed \eqref{eq:joint-throughput}. Compactness of the
feasible set gives an optimal solution whenever it is nonempty.

Given an optimal $(z,\xi)$, refresh whenever $N_k<C$. When $N_k\ge C$
and $S_k=s$, refresh with probability
\begin{equation}
\label{eq:joint-attainment}
b_s=\begin{cases}\xi_s/(pz_s),&z_s>0,\\0,&z_s=0.
\end{cases}
\end{equation}
Forward $\min\{N_k,C-X_k\}$ eligible events. These actions are causal
and physically feasible by the reduction lemma. Their transition matrix
has probability at least $1-p>0$ of jumping to state zero from every state.
Coupling chains at their first such reset proves uniqueness of the invariant
distribution and geometric convergence from any initial state. The balance
equations make $z$ that invariant distribution, with busy-refresh masses
$\xi$. Hence the stated cost and reward are attained, including at equality
of the QoS constraints.

Every state has cost at least $\Phi_A$; always refreshing attains that cost,
with $z_0=1$ and $\xi_0=p$. Thus the program is feasible exactly when
$\varepsilon\ge\Phi_A$. The cohort identity gives \eqref{eq:joint-exact},
and $\nu=m\lambda$ with $H_C=\nu-G_C$ gives \eqref{eq:joint-miss}.
At zero event rate the same isolated freshness floor is attained by
refreshing every cycle.
\end{IEEEproof}

\subsection{Proof of Corollary~\ref{cor:joint-small-A}}

\begin{IEEEproof}
Here $L=1$, $\omega_A(0)=\Phi_A$, and $\omega_A(1)=1$. Write
$c=\xi_0+\xi_1$. The balance equations give $z_0=1-p+c$, so the cost is
\[
\Phi_Az_0+1-z_0=1-(1-\Phi_A)(1-p+c).
\]
The budget requires $c\ge p-\delta_A$, with $c\ge0$.
This lower bound is attainable by a state-independent busy-refresh
probability $[p-\delta_A]^+/p$. Therefore
$\mathcal C_A^*=[p-\delta_A]^+$, proving \eqref{eq:joint-small-A} and
its specialization \eqref{eq:joint-Pplus1}.
\end{IEEEproof}

\subsection{Proof of Corollary~\ref{cor:joint-zero-stale}}

\begin{IEEEproof}
The zero-cost states are exactly $0,\ldots,R-1$. A feasible occupation measure
puts no mass elsewhere. The balance constraints give $z_s\le pz_{s-1}$
for $1\le s<R$, hence
\[
1=\sum_{s=0}^{R-1}z_s
\le z_0\sum_{s=0}^{R-1}p^s.
\]
Since $\sum_s\xi_s=z_0-(1-p)$, every feasible objective is at least
\[
\frac1{\sum_{s=0}^{R-1}p^s}-(1-p)
=\frac{p^R}{\sum_{s=0}^{R-1}p^s}.
\]
Refresh freely whenever $N_k<C$ and refresh in a busy cycle only at state
$R-1$. Then $z_s=p^s/\sum_{j=0}^{R-1}p^j$ for $0\le s<R$, and the
bound is attained with zero stale cost.
\end{IEEEproof}

\subsection{Proof of Proposition~\ref{prop:joint-provisioning}}

\begin{IEEEproof}
Additional service capacity, feeder capacity, or storage can be left unused,
so the optimal event rate is nondecreasing in $C$. Similarly, increasing $A$
weakly reduces every policy's stale fraction. This justifies
\eqref{eq:joint-provision} and the threshold comparison in
\eqref{eq:joint-Arequired}. Finite $C$ gives $G_C(\nu)<\nu$, so
$M_A^*>q$. Conversely, when $\varepsilon\ge\Phi_A$, refreshing every cycle
is freshness-feasible and supports $G_{C-1}(\nu)\to\nu$ as $C\to\infty$.
Thus $\eta>q$ is precisely the condition for finite required resources.

To prove downward closure in event load, start with any feasible policy
at $\lambda_2>\lambda_1>0$. At load $\lambda_1$, internally add independent
Poisson arrivals of rate $\lambda_2-\lambda_1$ and emulate that policy while
ignoring which packets are virtual. Selected virtual packets simply leave
their reserved resources idle. Conditional on the superposed arrival process,
each event is real with probability $\lambda_1/\lambda_2$, independently
of the emulated decisions. Expected real successes therefore scale by this
factor, preserving the event success fraction and the status decisions.
All virtual arrivals needed for a decision can be generated causally at the
corresponding contact. Thus feasibility at $\lambda_2$ implies feasibility
at $\lambda_1$.
\end{IEEEproof}

\bibliographystyle{IEEEtran}
\bibliography{ref}

\end{document}